# Electronic Structure, Magnetism and Superconductivity of Layered Iron Compounds

David J. Singh[a], Mao-Hua Du[a], Lijun Zhang[a], Alaska Subedi[a,b], Jiming An[a,c]

[a]Materials Science and Technology Division, Oak Ridge National Laboratory, Oak Ridge, TN 37830, U.S.A.
[b]Department of Physics and Astronomy, University of Tennessee, Knoxville, TN 37996, U.S.A.
[c]Wuhan University of Technology, Wuhan, China.

## Abstract

The layered iron superconductors are discussed using electronic structure calculations. The four families of compounds discovered so far, including Fe(Se,Te) have closely related electronic structures. The Fermi surface consists of disconnected hole and electron cylinders and additional hole sections that depend on the specific material. This places the materials in proximity to itinerant magnetism, both due to the high density of states and due to nesting. Comparison of density functional results and experiment provides strong evidence for itinerant spin fluctuations, which are discussed in relation to superconductivity. It is proposed that the intermediate phase between the structural transition and the SDW transition in the oxy-pnictides is a nematic phase.

David J. Singh
Materials Science and Technology Division
Oak Ridge National Laboratory
Oak Ridge, TN 37831-6114
U.S.A.
e-mail: singhdj@ornl.gov; Tel: +1-865-241-1944; FAX: +1-865-574-7659

## 1. Introduction

The layered iron superconductors (FeSC) [1] with critical temperatures now exceeding 50K provide a new window into high temperature superconductivity. This discovery raises fundamental questions in addition to interest in exploring this family to find and characterize the various superconducting compositions. In particular, it is important to establish the mechanism of superconductivity and the relationship to the other known family of high temperature superconductors, i.e. the cuprates. Here we discuss these issues using first principles calculations in relation to experiment.

## 2. Chemical Aspects

Superconductivity has now been discovered in several apparently related compounds. These include Fe-based oxypnictides (prototype $LaFeAsO_{1-x}F_x$) [1], $ThCr_2Si_2$ structure compounds ($BaFe_2As_2$) [2], LiFeAs [3] and PbO structure $Fe_{1+x}$(Se,Te) [4]. There are also structurally similar Ni compounds with low $T_c$, e.g. $BaNi_2As_2$ and LaNiPO [5,6], however the properties and superconductivity of these phases appear to be rather different from those of the Fe-based materials. In particular, the Ni phases can be understood within the context of conventional electron-phonon theory, [7] while the high $T_c$ FeSC are not described in this way [8,9].

The calculated local density approximation (LDA) density of states (DOS) of LaFeAsO is shown in Fig. 1, following Ref. [10]. As may be seen from the projections, the states within ~2 eV of the Fermi energy, $E_F$ are dominated by Fe d character. The As p states occur well below $E_F$, specifically below -2 eV, while the lower valence bands are derived from O p states.

Importantly, there is only modest hybridization between As p states and Fe d states, comparable to what often is found in oxides. Calculations for the Ni compounds and for $BaCo_2As_2$ show a similar structure, but we find rather more hybridization between Mn d states and As p states in $BaMn_2As_2$ and $BaMn_2Sb_2$. Returning to LaFeAsO, the DOS shows a pseudogap at an electron count of six per Fe. This is the same as the pseudogap in the DOS of bcc Cr and hypothetical non-spin-polarized bcc Fe. It is not, however, the position (four electrons) that would correspond to a tetrahedral crystal field. Thus, the basic chemistry of these compounds may be viewed as metallic sheets of $Fe^{2+}$ within an ionic background. This is consistent with the observation of high $T_c$ superconductivity in the non-pnictide $Fe_{1+x}(Se,Te)$.

## 3. Band Structure and Fermi Surface

The band structures of Fe-based superconductors have been reported by several authors [10-16]. A characteristic feature is a semi-metallic electronic structure, with relatively small disconnected electron and hole Fermi surfaces. In particular, these compounds generally show two intersecting electron cylinders centered at the zone corner compensated by hole cylinders at the zone center, and depending on the compound an additional three dimensional heavy hole section at the zone center (Fig. 2). The hole cylinders derive from bands associated with the lower d manifold (see Fig. 1), while the somewhat lighter mass electron cylinders come from bands dispersing downwards from the upper manifold. These basic features of the electronic structure are supported by spectroscopic measurements. [17,18] Importantly, it has been shown that these materials exhibit two gap superconductivity based on measurements of the upper critical field [19].

This strongly suggests that both the electron and hole Fermi surface sections are involved in superconductivity.

## 4. Magnetism

Despite the low carrier densities, these materials have relatively high density of states at the Fermi energy, $N(E_F)$ – high enough to place them near itinerant magnetic instabilities. Furthermore, the compensating electron and hole cylinders are nearly nested. They would be perfectly nested for equal size, equal shape Fermi surfaces. This nesting leads to nearness to a spin-density wave (SDW) instability, i.e. an instability towards itinerant antiferromagnetism at the nesting vector, which in 2D is (1/2,1/2) [9]. In fact, this SDW condenses, leading to a antiferromagnetic ground states for the undoped materials, with superconductivity appearing when the SDW is destroyed by doping with either electrons or holes or by pressure [20].

It is useful to discuss the magnetism starting from the bare Lindhard susceptibility, $\chi_0(\mathbf{q},\omega)$. The actual susceptibility is approximately given by an enhanced Lindhard form, $\chi(\mathbf{q},\omega) = \chi_0(\mathbf{q},\omega)[1 - I(\mathbf{q})\chi_0(\mathbf{q},\omega)]^{-1}$, which is the same as the Stoner formula for $\mathbf{q}=0$, $\omega=0$. As mentioned, the high $N(E_F)$ places these materials near the Stoner criterion for magnetism, even without nesting. The nesting yields on top of this a peak in $\chi_0(\mathbf{q},\omega)$ at the zone corner. As mentioned, for perfectly nested cylinders this diverges at the nesting vector, while variations in the shape will generally preserve a peak at the nesting vector but will smear the divergence. If the cylinders become mismatched in size, suppose with radii differing by $\delta q$, then the top will exhibit a flat area of diameter $2\delta q$, as sketched in Fig. 3. Eventually, if the size mismatch is large enough a depression at the nesting vector will appear [16] and at that point the SDW upon condensation will occur not exactly at the (1/2/,1/2) vector but will be shifted away from it to an incommensurate position. Fig. 3 also shows a schematic phase diagram based on the behavior of the Lindhard function, and assuming that the superconductivity is derived from itinerant spin fluctuations associated with the SDW ordering as in Ref. [9] (see below).

**5. Evidence for Strong Spin Fluctuations**

The magnetic and other properties of most itinerant systems are well described within the LDA, and often these can be further improved using more sophisticated density functionals, in particular generalized gradient approximations (GGAs). For example, the calculated magnetization per atom of bcc Fe is 2.19 $\mu_B$, 2.13 $\mu_B$ and 2.12 $\mu_B$ in the LDA, GGA and experiment, respectively [21]. On the other hand, the LDA and GGA often fail in strongly Coulomb correlated materials, such as the undoped Mott-Hubbard insulating cuprates. The nature of this failure is that by using a mean field like description of the Coulomb correlations, the tendency towards localization is underestimated. This leads to an underestimated tendency towards moment formation in the LDA. As a result the LDA predicts a paramagnetic metallic state for the undoped cuprates, while in fact the materials are antiferromagnetic insulators. In addition there is another, much smaller class of materials for which the LDA description of magnetism is grossly inaccurate. These are materials near magnetic quantum critical points. In this case the magnetic properties are renormalized by quantum fluctuations, which are beyond the scope of mean field like theories such as the LDA and GGA. These fluctuations lead to a reduced tendency towards magnetism, opposite to the effect of strong Coulomb correlations. The size of these renormalizations are strongly dependent on the material, and occur when there is a large phase space for competing spin fluctuations [22,23]. The signature of such physics is a large overestimate of the magnetic moments in the LDA and GGA as compared with experiment.

This is apparently the case in the FeSC. Moreover there are further signatures of strong spin fluctuations in the structural and spectroscopic data. Specifically, spin polarized GGA calculations done with optimized atomic positions (i.e. the GGA predicted structure), yield a structure, and in particular an As position in accord with experiment, but moments that are very much larger than experiment (~2 $\mu_B$ vs. ~0.4 $\mu_B$). Calculations without spin polarization on the other hand yield an As position closer to the Fe plane by more than 0.1 Å, which is an exceptionally large error for density functional calculations [24]. LDA magnetic calculations yield stable spin density wave states with lower moments but do not yield As positions in agreement with experiment. This is connected with a very large magnetoelastic coupling between the pnictogen position and Fe magnetism [24,25]. Furthermore, in the doped, non-magnetic superconducting compositions, one is again confronted with a conundrum. The As positions are again considerably higher than predicted in non-magnetic LDA and GGA calculations, similar to the case for the SDW phase, even though there is no ordered magnetism and additionally magnetic GGA calculations can again improve the As position but at the expense of having a stable high moment magnetic ground state. Thus it seems that there are strong quantum spin fluctuations in these materials that work against ordered magnetism. This is supported from the theory perspective by the fact that these materials have high density of states, which will lead to competition of the SDW with spin fluctuations at other **q**.

Experimental evidence includes transport data, which shows strong scattering above the SDW ordering temperature, susceptibility data showing an increasing $\chi(T)$ above the ordering temperature [26], and spectroscopic data showing large exchange splitting on short time scales [27] (see also Ref. 28). Thus it seems that there are strong itinerant spin fluctuations that modify the properties of both the SDW and superconducting phases of the layered Fe compounds.

**6. Superconducting State**

The superconducting state and mechanism have not yet been established in the FeSC. However, the rather different structure of the Fermi surface suggests that these materials could be different from the $d(x^2-y^2)$ state of the cuprates. In fact, opposite to the cuprates, experimental evidence presently available implies a node-less pairing state [29]. Since the electron-phonon interaction is apparently excluded based on first principles calculations [8,9], we discuss what might be expected based on spin-fluctuation as a pairing mechanism following Ref. [9]. Within Eliashberg theory, spin fluctuation pairing may be calculated using a pairing interaction that is closely related to the real part of the susceptibility. Importantly, this interaction has a negative (repulsive) sign for singlet pairing and a positive sign for triplet pairing [30]. Based on this, Mazin and co-workers [9] proposed an s-symmetry superconducting state where the electron and hole Fermi surfaces have opposite signs of the order parameter, i.e. an s(+/-) state. This state was also proposed by Kuroki and co-workers, with related considerations [31].

In fact, a node-less state is rather natural for a material with small Fermi surfaces. The reason is that stabilizing a state with nodes (sign changes) on a small Fermi surface would require a pairing interaction with an exceptionally strong **q-**dependence. The s(+/-) state, which involves only an inter-band sign change, is naturally stabilized by spin fluctuations associated with the (1/2,1/2) nesting of the electron and hole Fermi surfaces – the same nesting and interaction that leads to the SDW. While this state could be destroyed by inter-band scattering, it would be robust against low momentum transfer scattering. This in combination with the short coherence lengths may underlie the observed robustness of superconductivity against alloying with Co [32].

**7. Phase Diagram**

The experimental FeSC phase diagrams show an SDW, which gives way to superconductivity when destroyed either by doping or pressure. The evolution from an antiferromagnet to a superconductor with doping is superficially similar to the cuprates. However, the actual situation appears to be very different. Magnetism in cuprates is associated with a Mott state that has a substantial gap and is separated from the superconducting metal by a first order transition. In the FeSC, the magnetism appears to be itinerant in nature, and the SDW state retains a low density of metallic carriers. Thus the magnetic state and the superconductor are more clearly related. Furthermore the metallic cuprates, at least in the optimal to over-doped regime show large Fermi surfaces, while the FeSC show small disconnected Fermi surfaces – though with higher density of states. Finally, the band structures of the FeSC show less anisotropy than the cuprates, which may be important for applications.

Within a spin-fluctuation pairing scenario, as above, a qualitative phase diagram as sketched in the left panel of Fig. 3 results. The center of the phase diagram shows an SDW arising from a divergence of the real part of $\chi(\mathbf{q},0)$ due to inter-band nesting. Because there are often additional hole pockets and because the electron and hole cylinders are not perfectly nested, the best SDW ordering will be off-set from zero doping depending on the specific compound. As mentioned, superconductivity is associated with an integral over a region comparable in size to the Fermi surface size, while the SDW instability is associated with the peak value of $\chi$. As a result superconductivity will be favored as the electron and hole Fermi surfaces become mis-matched due to doping. However, since superconductivity and the SDW compete for the same Fermi surface and are driven by the same interactions, they will have at best only a small range of co-existence. Finally, it is interesting to note that within this scenario, small **q** scattering will be more destructive to the SDW than to superconductivity, even though s(+/-) is an unconventional superconducting state. This suggests that it may be possible to go from an SDW state to a superconducting state by introducing disorder into the materials, especially disorder that produces primarily low-**q** scattering.

One intriguing observation is that there is a double phase transition in undoped oxy-arsenides – first a structural transition to an orthorhombic or related symmetry compatible with the SDW, and then the SDW onset as the temperature is lowered [20]. This state is apparently not associated with a structural instability of the material independent of magnetism since it is not seen in superconducting samples, nor is there any unstable phonon in non-magnetic LDA calculations [8,10]. Instead the distortion is compatible with the symmetry of the SDW state although it occurs above the SDW ordering temperature. As mentioned, the SDW arises from condensation of spin-fluctuations at the M point (1/2,1/2). This leads to a structure in which there are chains of parallel spin Fe and these chains are arranged antiferromagnetically in the perpendicular direction, i.e. ferromagnetic along (1/2,-1/2) and antiferromagnetic due to the nesting along (1/2,1/2). Because of the origin of the SDW in the Fermi surface nesting, this will invariably lead to a highly anisotropic remaining Fermi surface, though the details will depend strongly on the value of the moments, which as mentioned is beyond the LDA. However, what is clear is that the spin wave dispersions in the SDW state will have a strong in-plane orthorhombic anisotropy. This is a characteristic of an itinerant nesting driven SDW, and is not expected for superexchange driven antiferromagnetism describable by a local Heisenberg model. In any case, it means that there may be a much shorter coherence length in one direction than in the orthogonal direction.

Considering now the melting of the SDW state with increasing temperature, one notes that this strong anisotropy, which should arise from the itinerant origin of the SDW, if strong enough can lead to an intermediate nematic phase. Other materials thought to have such a phase are $Sr_3Ru_2O_7$ near the field induced metamagnetic quantum critical point [33] and cubic MnSi, which is also near a quantum critical point associated with itinerant magnetism [34]. Such a phase would be consistent with the observed behavior of the intermediate phase – transport, and other Fermi surface dependent properties as well as orthorhombic distortion similar to the SDW but no long range 3D magnetic order. A key feature of such phases is that they have partial order, but retain a large phase space for fluctuations, which is essential for stabilizing them. This is different from nano-scale domains, for example, which would not have large phase spaces for fluctuations. A nematic state should be detectable in magnetic diffuse scattering for example. Another key experiment will be mapping of the imaginary part of $\chi(\mathbf{q},\omega)$ in the Brillouin zone, e.g. with neutron scattering. This provides a test for the anisotropy of the spin wave dispersions as well as providing confirmation of the itinerant nature of the spin fluctuations, e.g. through the **q** dependence and its relationship with the Fermi surface. We note that a nematic intermediate phase was proposed by Fang and co-workers, based on a different model [35]

## 8. Summary and Conclusions

The properties of the FeSC have been discussed from the point of view of electronic structure calculations. The basic picture that emerges is of low carrier density metals with separated Fermi surfaces and proximity to itinerant magnetism, including an itinerant SDW. This is very different from the situation in the cuprates. As such the FeSC provide an entirely new class of high $T_c$ superconductors, distinct from electron-phonon superconductors and also from the cuprates.

**Acknowledgements**

We are grateful for discussions with I.I. Mazin, M.D. Johannes, B.C. Sales, D. Mandrus, A.S. Sefat and M.A. McGuire. This work was supported by the U.S. Department of Energy, Division of Materials Sciences and Engineering.

**Figure Captions**

Figure 1. Calculated DOS and projections of LaFeAsO following Ref. [10]. Note that the Fermi energy lies at the bottom of a pseudogap and that the states near the Fermi energy have Fe d character. Note that anion wavefunctions extend well beyond the sphere radii used in the projections, and therefore the anion projections are proportional to but underestimate the anion contributions.

Figure 2. Fermi surface of LaFeAsO following Ref. [10]. The arrow indicates the approximate nesting of the hole and electron sheets.

Figure 3. The left panel shows the schematic evolution of $\chi_0(\mathbf{q},0)$ with electron doping, with the region shown corresponding to the region around the nesting vector (1/2,1/2). The right panel shows a schematic phase diagram based on this (see text).

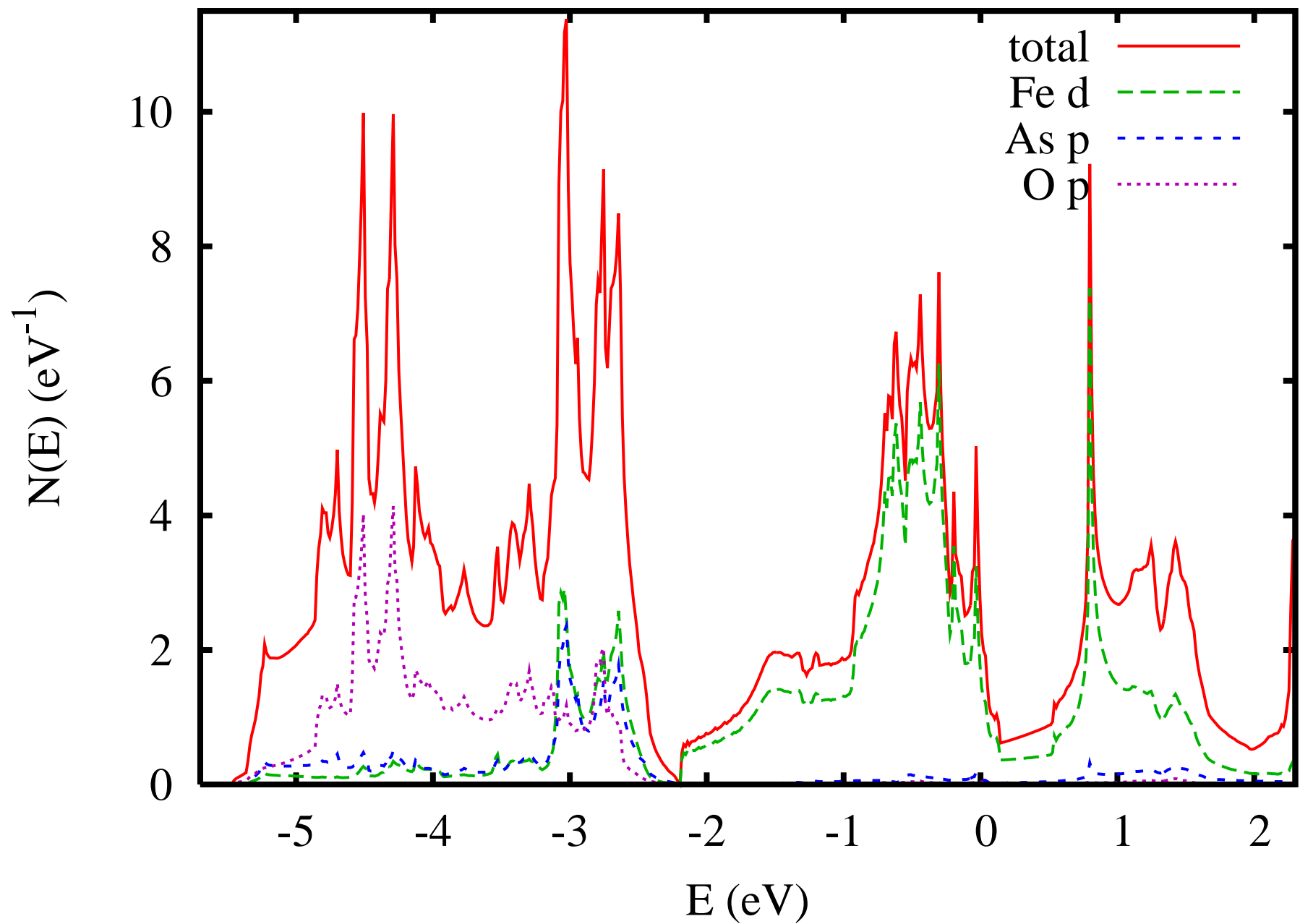

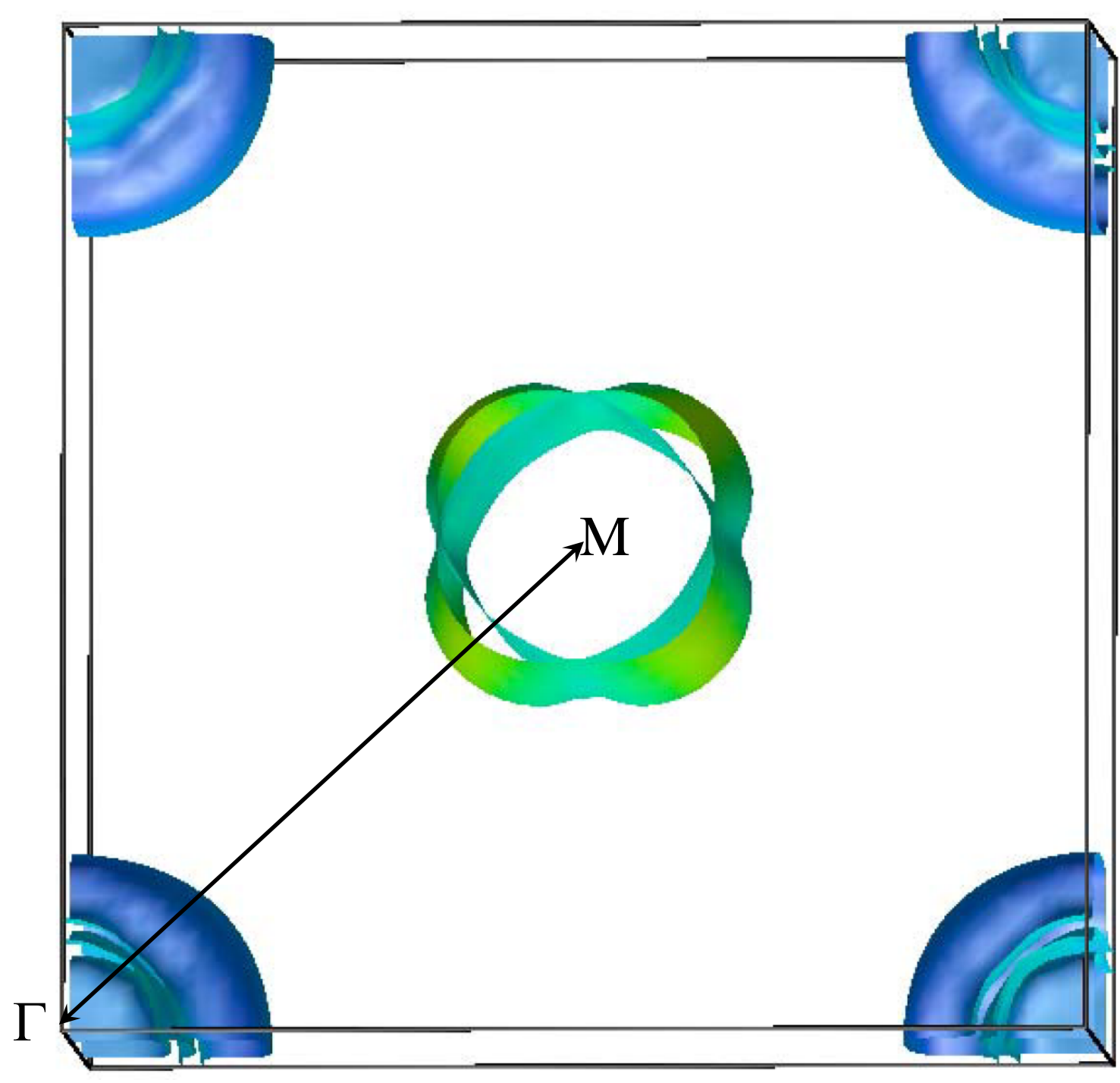

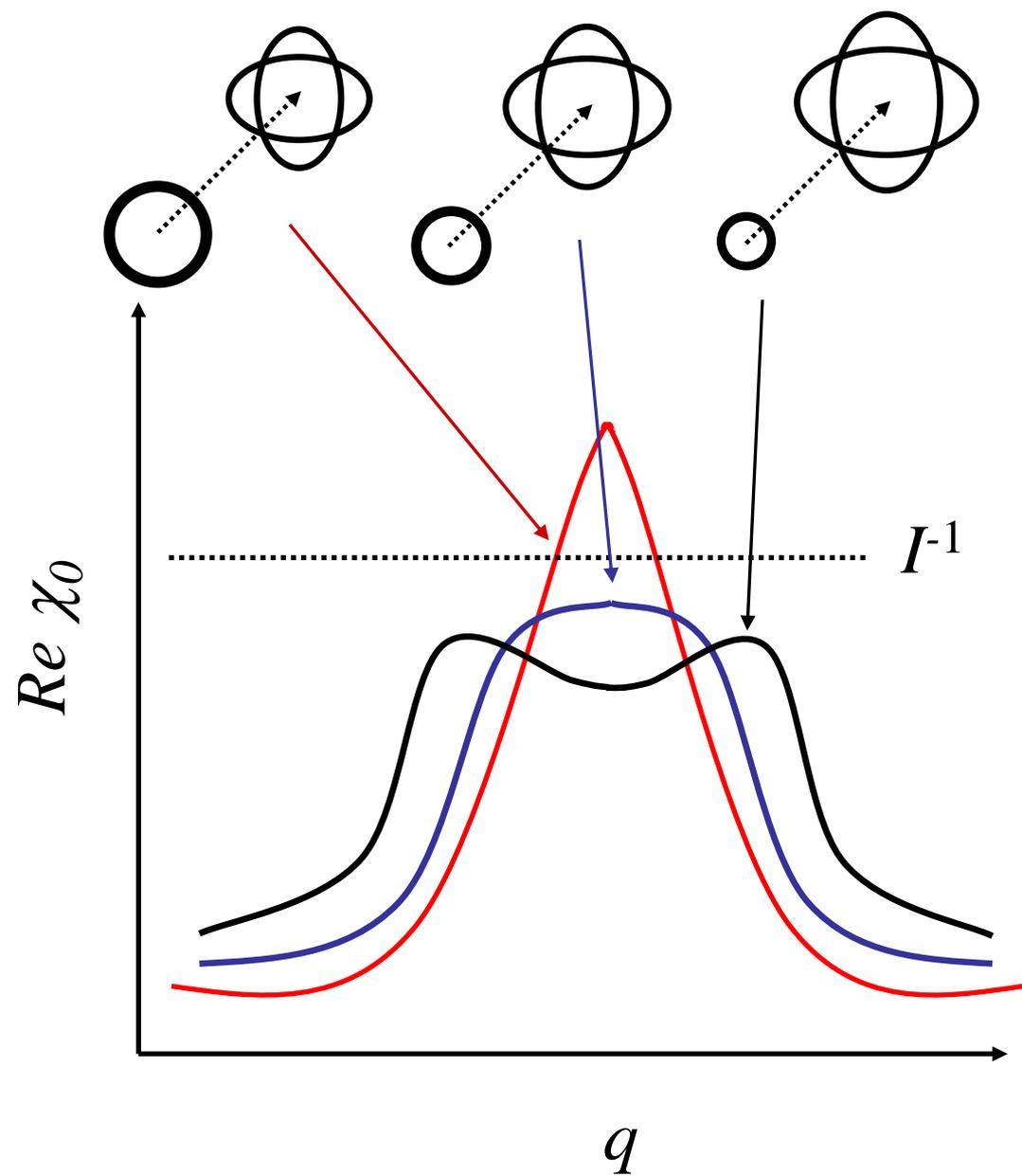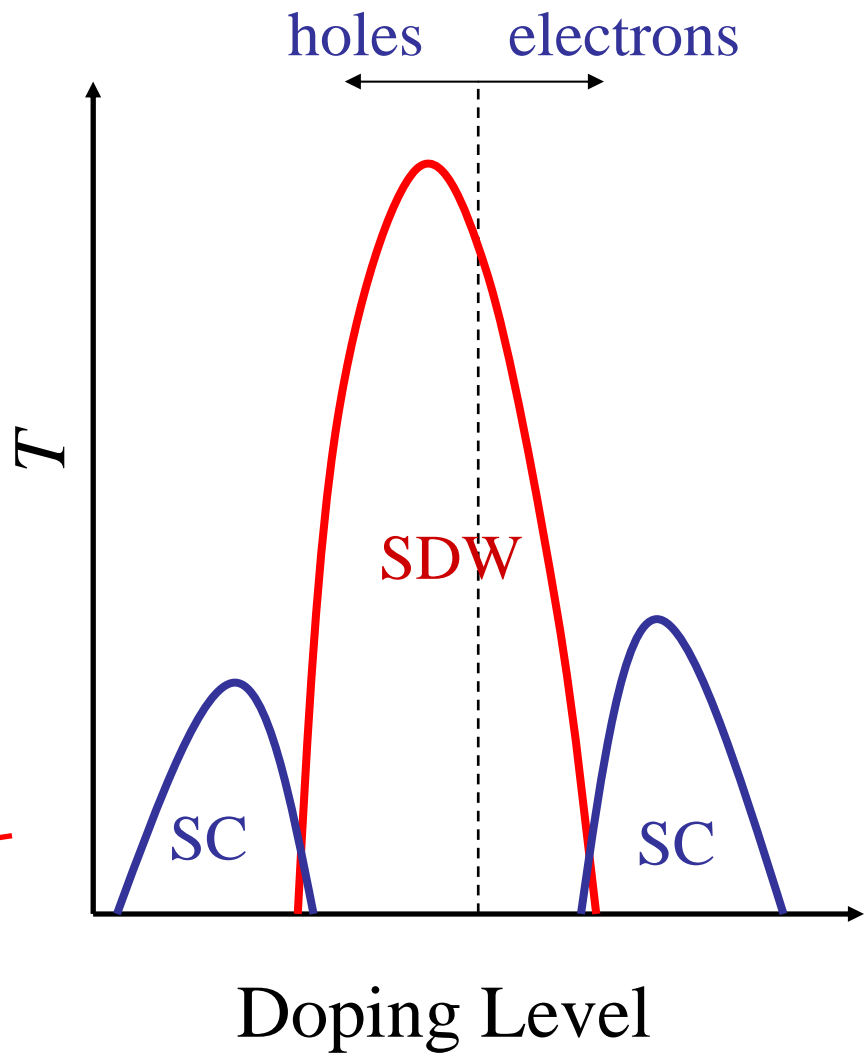